\documentclass[10pt,journal,letterpaper,compsoc]{IEEEtran}

\usepackage{mathptmx}
\usepackage{wrapfig}
\usepackage{graphicx}
\usepackage{times}
\usepackage{color}
\usepackage{soul}
\usepackage{wrapfig}
\usepackage{picins}
\usepackage{caption3}
\usepackage{multirow}
\usepackage{booktabs}
\usepackage[bookmarks,backref=true,linkcolor=black]{hyperref} %,colorlinks
\hypersetup{
  pdfauthor = {},
  pdftitle = {},
  pdfsubject = {},
  pdfkeywords = {},
  colorlinks=true,
  linkcolor= black,
  citecolor= black,
  pageanchor=true,
  urlcolor = black,
  plainpages = false,
  linktocpage
}

\begin{document}

\title{\Title}
\title{Curve Networks for Surface Reconstruction}
\author{Yuanhao Cao,
        Liangliang Nan$^*$\thanks{$^*$ Corresponding author.},
        Peter Wonka

\IEEEcompsocitemizethanks{
	\IEEEcompsocthanksitem The authors are with the Visual Computing Center, King Abdullah University of Science and Technology, Thuwal 23955-6900, Saudi Arabia.\protect \\
	E-mail: ppxhappy@126.com, \{liangliang.nan, pwonka\}@gmail.com.}
}

\IEEEcompsoctitleabstractindextext{
\begin{abstract}
 Man-made objects usually exhibit descriptive curved features (i.e., curve networks). The curve network of an object conveys its high-level geometric and topological structure. We present a framework for extracting feature curve networks from unstructured point cloud data. Our framework first generates a set of initial curved segments fitting highly curved regions. We then optimize these curved segments to respect both data fitting and structural regularities. Finally, the optimized curved segments are extended and connected into curve networks using a clustering method. To facilitate effectiveness in case of severe missing data and to resolve ambiguities, we develop a user interface for completing the curve networks. Experiments on various imperfect point cloud data validate the effectiveness of our curve network extraction framework. We demonstrate the usefulness of the extracted curve networks for surface reconstruction from incomplete point clouds.
\end{abstract}

% Note that keywords are not normally used for peer review papers.
}

\maketitle
\IEEEdisplaynotcompsoctitleabstractindextext
\IEEEpeerreviewmaketitle

%% if specified like this the section will be ommitted in review mode
\section{Introduction}
\label{sec:intro}
\begin{figure*}[th]
	\centering
	\includegraphics[width=1.0\linewidth]{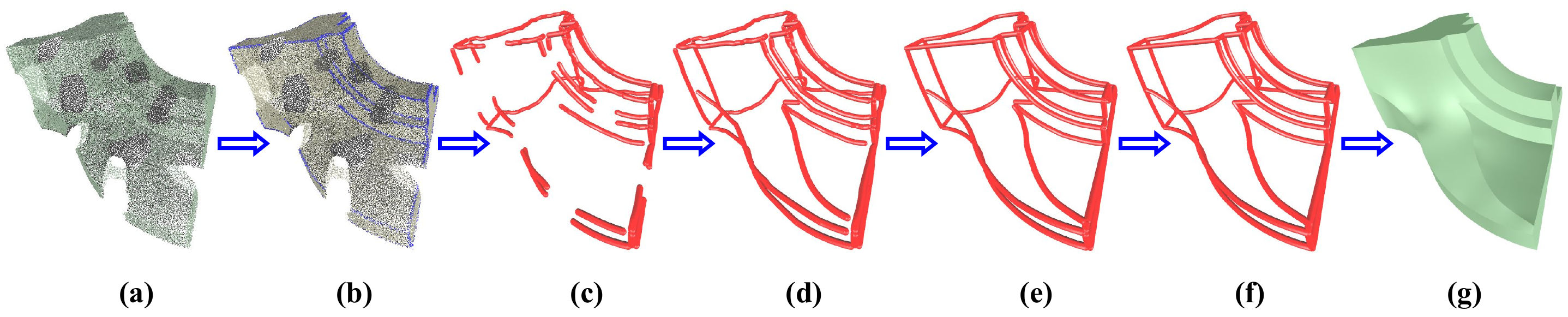}
	\caption{Algorithm overview. (a) Input point cloud; (b) detected feature points;
		(c) initial curve segments; (d) improved feature curves with user guidance; (e) optimized feature curves; (f) final curve networks after completion; (g) the reconstructed surface model from the curve networks using the techniques of~\cite{Zhuang2013}.}
	\label{fig:pipeline}
\end{figure*}

Man-made objects usually exhibit descriptive curved features. These curved features, we call \emph{curve networks}, convey the structural characteristics of the objects. From the visual perception perspective, they serve as high-level representation of the objects. Various applications, such as non-photorealistic rendering~\cite{Saito1990}, product design~\cite{True2Form2014}, abstraction~\cite{Mehra2009}, segmentation~\cite{Nieser2010}, reconstruction~\cite{abbasinejad2011}, and shape editing~\cite{gal2009iwires}, have exploited and benefited from the extracted curve networks. In this work, we are interested in the problem of extracting curve networks from noisy and incomplete point clouds and using these extracted networks to guide surface reconstruction from incomplete point clouds.

In the last decades, the prevalence of various laser scanners and depth cameras (e.g., Kinect) enabled non-professional users to obtain a 3D sampling of an object in a matter of seconds. However, due to occlusions and specific material properties (e.g., transparent, reflective), obtaining a point cloud with reasonably coverage of an object remains a challenge. In practice, it is quite common that significant portions of the object are either under-sampled or completely missing in a 3D point cloud. This limits the applicability of the widely available acquisition devices and hinders the application of the large amount of existing point cloud data.

The main problem for reconstruction from partial point cloud data comes from the lack of constraints in the missing regions. Thus, the reconstruction is ill-posed as an infinite number of valid surfaces may pass these regions. To compensate for the lack of constraints, smoothness is usually exploited to fill the holes in the reconstructed surface models. However, the smoothness constraints are too local to fill holes occurring near sharp features. High-level constraints, such as symmetry, may provide an efficient completion tool~\cite{law2011single}. However, when symmetry is not applicable, or the data is highly incomplete, it is impossible to infer a faithful 3D completion and reconstruction. Given the descriptive characteristics of curve networks, the motivation of this work is to extract such curve networks from partial point cloud data and utilize them as geometric and topological constraints to regularize the ill-posed surface reconstruction problem. To this end, we present a hybrid framework for extracting high quality curve networks from unstructured point cloud data that may have severe missing regions.

Our overall contributions are as follows:
\begin{itemize}
	\item a novel framework that can effectively extract curve networks form partial point clouds.
	\item an optimization algorithm exploiting structural regularities to enhance the extracted curve networks to be regular and meanwhile respect the input point clouds.
	\item we demonstrate that the extracted curve networks can significantly regularize surface reconstruction from incomplete point clouds.
\end{itemize}

\section{Related Work}
\label{sec:related}
There exists a large volume of work related to surface reconstruction in literature. In this section, we mainly review the work that are closely related to feature detection, curve based modeling, surface reconstruction from curves, and curve network extraction.

\textbf{Feature detection}.
Quite a few techniques have been proposed for detecting curved features on polygonal models and point clouds.
Lee et al. propose geometric snake~\cite{Lee2002}, an interactive tool for detecting curved features from triangular meshes by extending the 2D active contour model (snakes) to 3D surfaces. The user sketches initial feature curves on the input surface, and the 3D snake iteratively snaps them to the curved features in the surface. Ohtake et al.~\cite{Ohtake2004} exploit implicit surface fitting to calculate extremal coefficients for extracting ridges and valleys from mesh surfaces. Kim et al.~\cite{KimK06a} utilize a variant of Moving-Least-Squares method to fit local surface patches in the neighborhood of each vertex, and then compute local curvatures based on the fitted local surface patches. Similarly, Yoshizawa et al.~\cite{Yoshizawa2005} extract crest lines by estimating the curvature tensor and curvature derivatives based on local polynomial fitting. In the work of Nomura and Hamada~\cite{Nomura2001}, the authors detect feature curves by calculating the skeleton of the feature region defined by the concavity and convexity.

To handle random noise, outliers, and artifacts, Min et al.~\cite{Park2012} present a method based on the tensor voting to extract sharp features from unstructured point clouds. Mark et al.~\cite{MarkPauly2003} extract feature lines from point-sampled geometry by computing a minimum spanning graph of feature nodes that have high probability belonging to a feature.
By fitting spline curves, Joel et al.~\cite{Joel2008} identify sharp features in a point-based model and align the spline curves  with the sharp edges of the model.

\textbf{Curve based modeling}.
There are also techniques and systems that are able to transfer 2D sketches into 3D representations. For example, ILoveSketch~\cite{Bae2008}, a 3D curve sketching system that captures some of the affordances of pen and paper for professional designers, allows a designer to iterate directly on conceptual 3D models. Other systems, such as Teddy~\cite{Igarashi1999} and Fibermesh~\cite{FiberMesh}, provide simple user interfaces for designing freeform surfaces from a collection of 2D sketches. The user first creates a rough 3D model by drawing 2D strokes.  Then the 3D surface models can be further edited by sketching directly on the models, where the 3D curves serve as handles for controlling the geometry. Recently, Baoxuan et al.~\cite{True2Form2014} design a sketch-based modeling system (i.e., True2Form) that reconstructs 3D curve networks from typical 2D design sketches. Their strategy relies on prior knowledge to enforce structural regularities of an object.

\textbf{Surfaces reconstruction from curves}. The motivation of this task is to recover full geometry from a set of curves spanning in the surfaces. Orbay and Kara~\cite{Orbay2011} propose a sketch-based modeling interface for creating smooth surfaces from curve networks. Based on a linear algebra representation of suface patches, Abbasinejad et al.~\cite{abbasinejad2011} introduce a system that supports automatic generation of piecewise smooth surfaces from curve networks. With similar motivation, Bessmeltsev et al.~\cite{Bessmeltsev2012} present a design-driven approach for quadrangulating closed 3D curve networks. Zou et al.~\cite{Zou2013} present an algorithm for triangulating 3D spatial polygons. To fill holes, an N-sided hole filling technique proposed by Tam{\'a}s et al.~\cite{varady2011transfinite} can interpolate the boundary curve of each hole. For partial scans with large missing parts, Nan et al.~\cite{nan20142d3d} propose to lift 2D image boundaries into 3D space to constrain the ill-posed surface reconstruction problem.

\textbf{Curve network extraction}.
To extract closed curve networks, Demarsin et al.~\cite{Kris2007} propose an algorithm for extracting closed sharp feature lines from point clouds. Based on first order segmentation, they first extract candidate feature points and then represent them as a graph to recover the sharp feature lines. Then a minimum spanning tree is constructed to enclose these curved lines. In the work of Cao et al.~\cite{Yuanhao2015}, the authors extract curve networks by first detecting curved segments and then extending them to closed curve loops on surfaces.

Following the work of~\cite{Yuanhao2015} and~\cite{True2Form2014}, we extract curve networks from partial and noisy point clouds through optimization. Our formulation enforces the detected curve networks to respect both data fitting and structural regularities of the objects.

\section{Overview}
\label{sec:overview}
Given a noisy and incomplete point cloud as input, our goal is to extract complete curve networks from such an imperfect input point cloud. Our framework consists of the following three key steps (an overview of our approach is shown in Fig.~\ref{fig:pipeline}):

\textbf{Curved segment generation}. We first compute the surface variation at each point and extract regions of high-level variation using simple thresholding. Then, initial curved segments are generated by fitting curves to those feature points in the point cloud. For partial point clouds, we develop a simple user interface allowing a user to guide the curve fitting through loosely sketching strokes on the 3D point clouds (Sec.~\ref{sec:initial_extraction}).

\textbf{Curved segment optimization}. We introduce an energy minimization formulation to optimize the feature curves. Our objective function is designed to enforce data fitting and the smoothness of the curved feature, and meanwhile to respect the structural regularities of the point cloud (Sec.~\ref{sec:optimization}).

\textbf{Curve network completion}. After the curved segments being optimized, we extend and connect these individual feature curves to generate complete curve networks. We use the algorithm proposed by Zhuang et al.~\cite{Zhuang2013} to detect cycles of surface patches from the curve networks (Sec.~\ref{sec:connection}).

\section{Method}
\label{sec:method}
Given a noisy and incomplete point cloud as input, we first identify feature points in the point cloud. Then, we extract curved segments from these feature points via curve fitting. We also allow users to guide the curve fitting through a simple user interface for large missing regions in the point cloud.

\subsection{Curved segment generation}
\label{sec:initial_extraction}
We first detect feature points as those having high surface variations in the point cloud. Specifically, we use the method described in~\cite{MarkPauly2003} to define the variation $\sigma(\textbf{p})$ at a point $\textbf{p}$ as
\begin{equation}\label{variation}
	\sigma(\textbf{p}) = \frac{\lambda_{1}}{\lambda_{1}+\lambda_{2}+\lambda_{3}},
\end{equation}
where $\lambda_{i} (1 \leq i \leq 3)$ denote the three eigenvalues (in an ascending order) of the covariance matrix  defined on the neighborhood of $\textbf{p}$. For more details on the covariance matrix setup, please refer to~\cite{MarkPauly2003}. Since $\sigma(\textbf{p})$ is invariant under scales at each point, so feature points can be identified by simply thresholding surface variations at each point in the input point cloud. Specifically, a point $\textbf{p}$ is considered as a feature point if $\sigma(\textbf{p}) > \sigma_t$, where $\sigma_t$ is the threshold that is set to 0.04 in our experiment.

\textbf{Curve segment extraction}.
From the identified feature points, we generate polylines using a modified version of~\cite{DanielsII2008} to fit the points with high variations.

% Given a maximum allowable segment length $s_{max}$ that is to control the resolution of the feature polylines, each polyline is initialized at a seed point $\textbf{p}$ in the feature region. Here we only consider feature points that fall in the neighborhood of $\textbf{p}$, i.e., $\textbf{Nh}_{s_{max}}\subset \textbf{N}_{s_{max}}$.
% We then use PCA (principle component analysis) to compute the extending direction for each polyline, which is the direction of the eigenvector $\textbf{v}_{t}$ corresponding to the largest eigenvalue of the covariance matric of $\textbf{Nh}_{s_{max}}$. Then all the points $\textbf{p}_{i}\in \textbf{Nh}_{s_{max}}$ are projected onto the line defined by the seed point $\textbf{p}$ and the vector $\textbf{v}_{t}$, yielding points $\textbf{p}^{\prime}_{i}$. The seed point $\textbf{p}$ now can be extended to two points $\textbf{p}_{j}$ and $\textbf{p}_{k}$ in $\textbf{Nh}_{s_{max}}$ with the furthest corresponding projections $\textbf{p}^{\prime}_{j}$ and $\textbf{p}^{\prime}_{k}$. This procedure is repeated at $\textbf{p}_{j}$ and $\textbf{p}_{k}$ until no points remain in the growth direction. The parameter $s_{max}$ is set to $d/200.0$ in our experiments, where $d$ is the diagonal length of the bounding box of the point cloud.

\begin{figure}[t]
	\centerline
	{
		\includegraphics[width=0.9\linewidth]{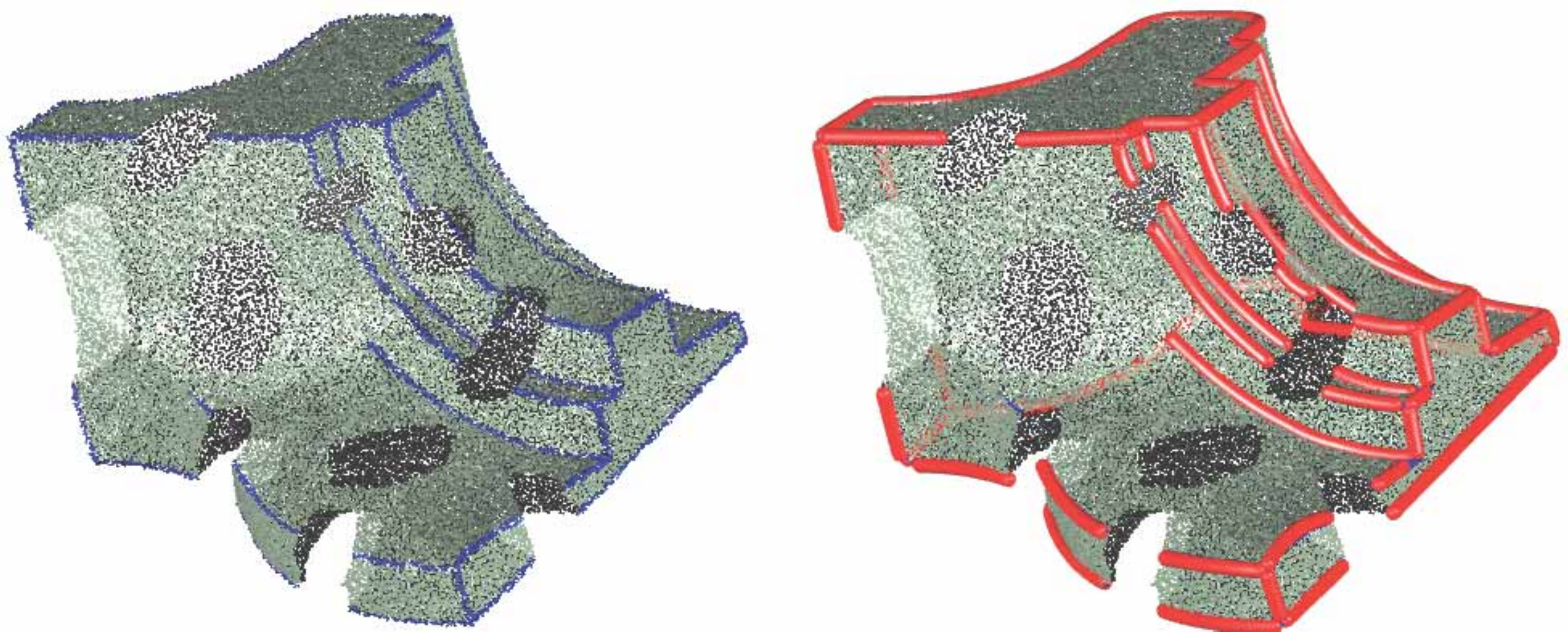}
	}
	\caption{Extracted feature points (left, in blue) and initial curved segments (right, in red) extracted from the point cloud.}
	\label{fig:variation}
\end{figure}

To handle noisy point clouds, we propose an additional termination condition to prevent the polyline growing into multiple feature regions at sharp corners. Given an endpoint $\textbf{p}_{k}$ and it direct neighbor $\textbf{p}_{k-1}$ in the polyline, we extend $\textbf{p}_{k}$ to $\textbf{p}_{k+1}$ if the angle between $\overline{\textbf{p}_{k-1}\textbf{p}_{k}}$ and $\overline{\textbf{p}_{k}\textbf{p}_{k+1}}$ is smaller than $30^{\circ}$. In our implementation, we sort the feature points according to their variations and choose the point with highest variation as the seed point for propagating a polyline. After one polyline propagation is terminated, all the feature points within the neighborhoods $\textbf{N}_{s_{max}}$ to the polyline are removed. We then choose a new seed point from the remaining feature points and propagate another polyline. We repeat this process until no feature points are left. Fig.~\ref{fig:variation} illustrate the feature points (left) and the extracted polylines (right) in the point cloud.

\begin{figure}[t]
	\centerline
	{
		\includegraphics[width=0.8\linewidth]{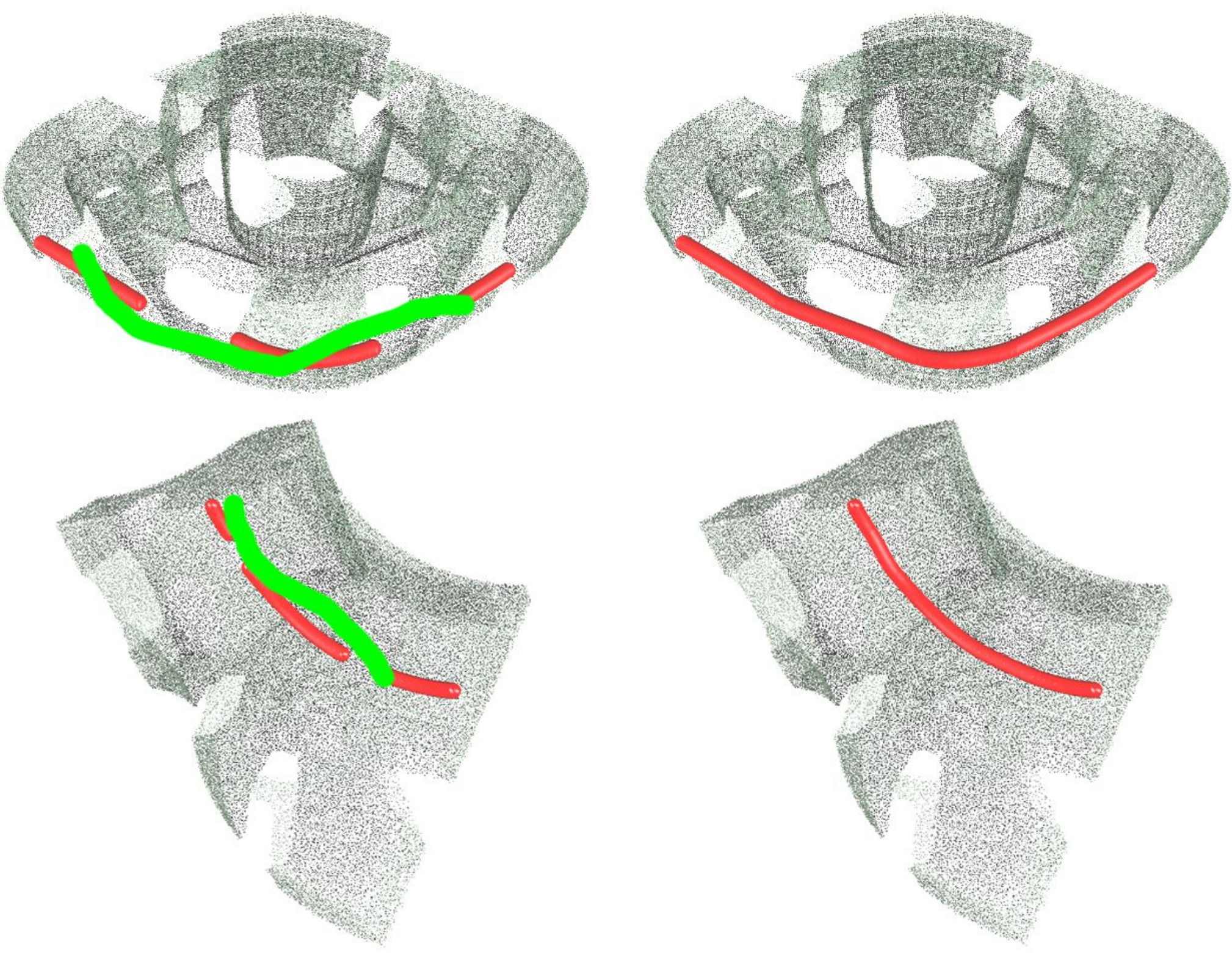}
	}
	\caption{Two discontinuous feature curves (left column) are smoothly connected by simple user strokes, yielding more complete feature curves (right column). Feature curves are in red and user strokes are in green.}
	\label{fig:interaction}
\end{figure}

\textbf{User guidance and symmetry}.
For point clouds with large missing regions, our automatic curve segment generation may fail to extract good polylines. So we develop a user interface that allows users to guide the feature curve extraction process by simple clicking and sketching directly on the partial 3D point clouds (please refer to the accompanying video). Fig.~\ref{fig:interaction} shows two examples of how users' guidance can help to obtain more complete and smooth feature curves.

\begin{figure}[t]
	\centerline
	{
		\includegraphics[width=0.8\linewidth]{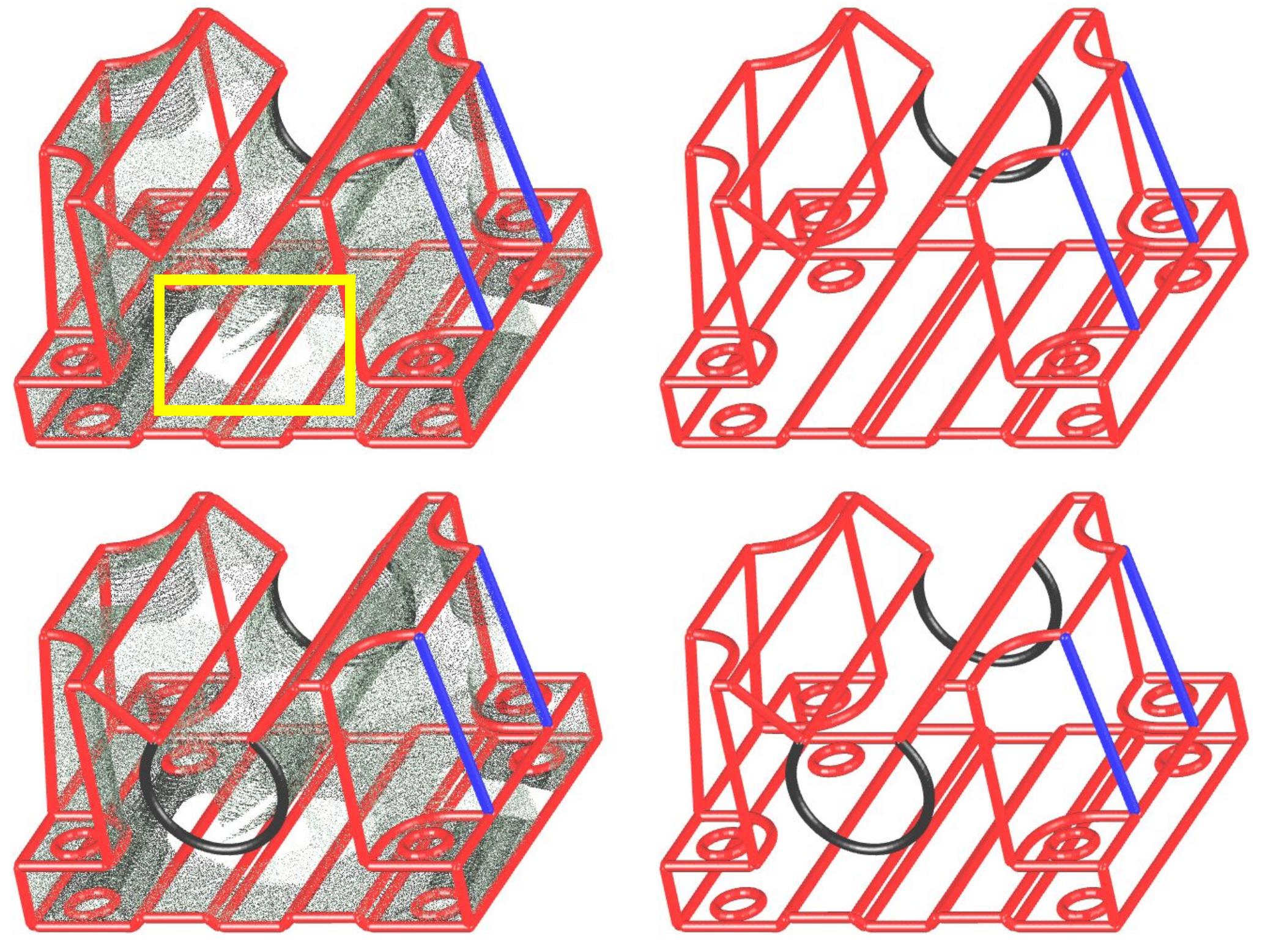}
	}
	\caption{A missing feature curve in the region marked by the yellow rectangle is generated from its mirror image (the black curve in the upper right subfigure) using reflective symmetry information. Left: feature curves overlaid on the point clouds. Right: feature curves only.}
	\label{fig:symmetry}
\end{figure}

Since reflective symmetry is common for man-made objects, we allow the user to indicate if symmetry is applicable for the objects represented by point clouds. For partial point clouds\textsl{}, automatic symmetry detection is usually not reliable. Thus, we rely on the extracted feature curves (even though they are not complete) to determine the symmetry plane which in turn completes the feature curves based on reflective symmetry. Fig.~\ref{fig:symmetry} shows an example of completing the initial feature curves by symmetry.

\subsection{Curved segment optimization}
\label{sec:optimization}
The initial curved segments extracted in the previous step have two problems. On the one hand, they are extracted from local sampled points, thus they are not accurate due to noise and outliers. On the other hand, they are not smooth enough for further reconstruction due to the nature of the polyline growing process. Thus, we propose to exploit prior knowledge and global structural regularities of the object to optimize the entire feature curves altogether.  Our optimization encourages obtaining smooth, globally regular feature curves, and meanwhile respecting the input point cloud.

Let $\mathbf{B}_{1}, \mathbf{B}_{2}, ... , \mathbf{B}_{m}$ be the $m$ initial curved segments, and each curve $\mathbf{B}_{i}$ is represented by $n_{i}$ sequential discrete points $\mathbf{b}_{i,1}, \mathbf{b}_{i,2}, ... , \mathbf{b}_{i, n_{i}}$.  The optimized position of points $\mathbf{b}_{i,j}$ is represented by $\overline{\mathbf{b}}_{i,j}$. We define the following energy terms for curve optimization:

\begin{itemize}
	
	\item \textbf{Data fitting}. We use the data fitting term to ensure the optimized curved segments stay as close as possible to the 3D feature points detected from the point cloud. Mathematically, this term is define as the sum of the squared distance between a point $\overline{\mathbf{b}}_{i,j}$ in the feature curve and the 3D point in the input point cloud:
	\begin{equation}\label{align}
		E_{alignment} = \sum_{i=1}^{m}\sum_{j=1}^{n_{i}}\|\overline{\mathbf{b}}_{i,j}-\mathbf{p}_{i,j}\|^{2},
	\end{equation}
	where $\mathbf{p}_{i,j}$ is the variation weighted average of all the neighboring points of $\mathbf{b}_{i,j}$.
	
	\item \textbf{Smoothness}. This term encourages the curved segments to deform to be smooth. We define the non-smoothness $F_{smooth}$ of the curved segments as following:
	\begin{equation}\label{smooth}
		E_{smooth} = \sum_{i=1}^{m}\sum_{j=2}^{n_{i}-1}\|\overline{\mathbf{b}}_{i,j-1}-2\overline{\mathbf{b}}_{i,j}+\overline{\mathbf{b}}_{i,j+1}\|^{2}.
	\end{equation}

	\item \textbf{Fidelity}. This term prevents the curved segments from deviating too much from their initial locations. It is defined as the sum of the squared distance of a point in the feature and its initial position:
	\begin{equation}\label{fidelity}
		E_{fidelity} = \sum_{i=1}^{m}\sum_{j=1}^{n_{i}}\|\overline{\mathbf{b}}_{i,j}-\mathbf{b}_{i,j}\|^{2}.
	\end{equation}
	
\end{itemize}

Then our objective function $E$ is defined as the weighted sum of the above individual energy terms:
\begin{equation}\label{energy_function}
	\begin{array}{l}
		E=\omega_{1} \cdot E_{fidelity}+\omega_{2} \cdot E_{alignment}+\omega_{3} \cdot E_{smooth}.
	\end{array}
\end{equation}

\begin{figure}[t]
	\centerline
	{
		\includegraphics[width=0.8\linewidth]{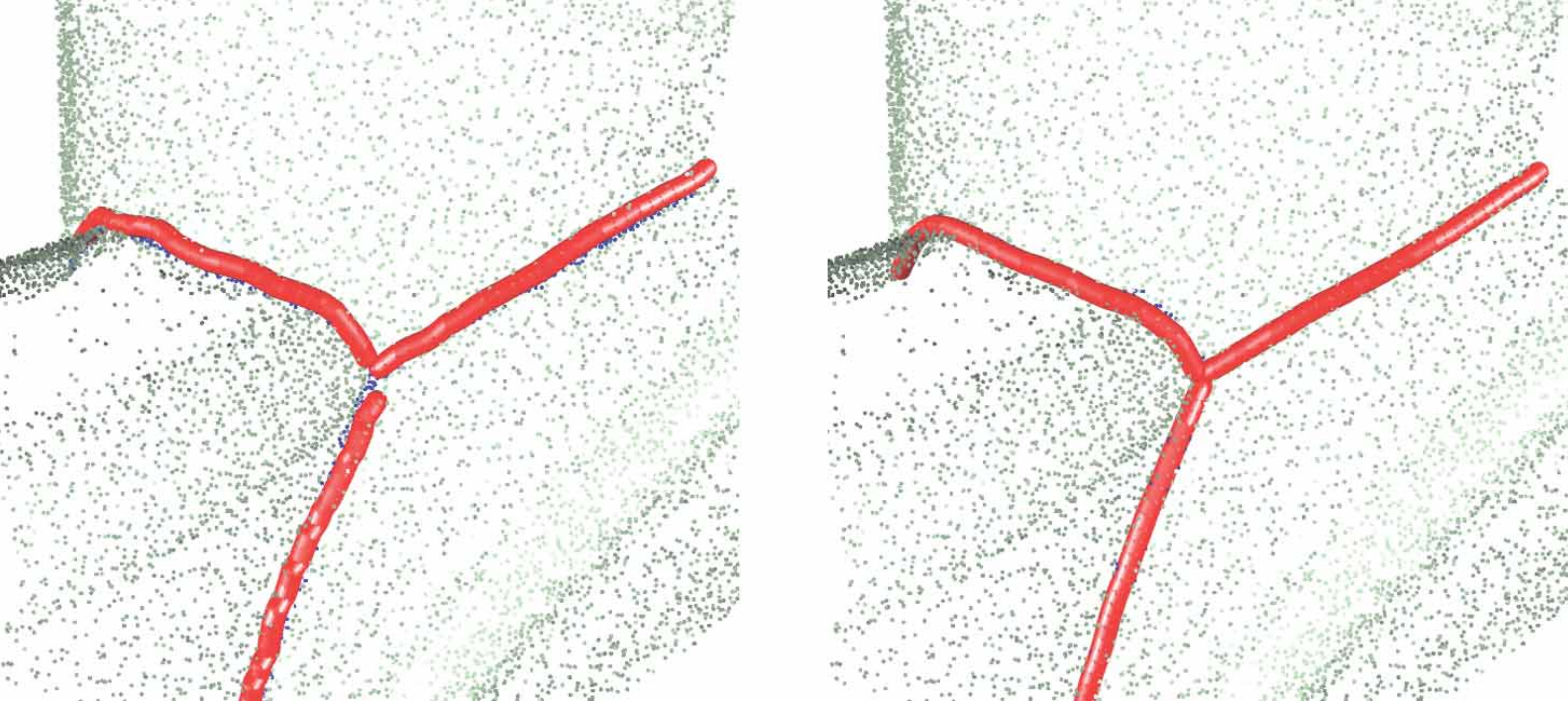}
	}
	\caption{Feature curve optimization without shape prior constraints. Left: The initial curved segments extracted in the previous step. Right: The optimized feature curves, which are smooth and are aligned well with the feature points (in blue).}
	\label{fig:align_smooth}
\end{figure}

By minimizing the above energy function, the quality of the initially detected curved segments are improved. Fig.~\ref{fig:align_smooth} shows an example of the optimization result without structural regularity constraints. As can be seen from this figure, the curved segments have been smoothed and are now better aligned with the detected 3D feature points in the point cloud.

We observe structural regularities (such as straight line segments, co-planarity, and symmetry) are common features in man-made objects. Thus, we exploit these high-level structural regularities to further improve the quality of the feature curves. In this work, the following structural regularities are detected and enhanced:

\begin{itemize}
	\item \textbf{Linearity}. To straighten feature curves representing straight line segments, this
	term measures how well the interior points of a curved segment can be represented by the linear combination of its two endpoints:
	\begin{equation}\label{line}
		E_{line} = \sum_{i\in \mathbf{S}_{line}} \sum_{j=1}^{n_{i}}\|t_{i,j} \cdot \overline{\mathbf{b}}_{i,1}+(1-t_{i,j}) \cdot \overline{\mathbf{b}}_{i,n_{i}}-\overline{\mathbf{b}}_{i,j}\|^{2},
	\end{equation}
	where $\mathbf{S}_{line}$ are the feature curves detected as straight lines. Parameter $t_{i,j}$ is the weight that can be computed by minimizing $\|t_{i,j} \cdot \mathbf{b}_{i,1}+(1-t_{i,j}) \cdot \mathbf{b}_{i,n_{i}}-\mathbf{b}_{i,j}\|^2$ in the initial feature curves.
	
	\item \textbf{Circularity}. For a closed curve $\mathbf{B}_{i}$ representing a circle, the circularity term is defined to measure how far the closed curve is from being a perfect circle. We measure the non-circularity as the sum of the difference between the squared length of the circle's diameter $r_{i}$ and the squared length of the segment from a point in the feature curve to the circle center $\mathbf{c}_{i}$:
	\begin{equation}\label{circle}
		E_{circle} = \sum_{i\in \mathbf{S}_{circle}}\sum_{j=1}^{n_{i}}(\|\overline{\mathbf{b}}_{i,j}-\mathbf{c}_{i}\|^{2}-r_{i}^{2}).
	\end{equation}
	where $\mathbf{S}_{circle}$ are the feature curves detected as circles.
	
	\item \textbf{Co-planarity}. For all the curved segments $\mathbf{S}_{coplanar}$ that are supposed to be lying in the same plane, we first compute a plane $\mathbf{C}$ by least-squares fitting of the feature points. Then this term is defined as the sum of the squared distance between a point in the feature point and the plane:
	\begin{equation}\label{coplanar}
		E_{coplanar} = \sum_{i\in \mathbf{S}_{coplanar}}\sum_{j=1}^{n_{i}}(dist(\overline{\mathbf{b}}_{i,j}, \mathbf{C}))^2,
	\end{equation}
	where $dist(\overline{\mathbf{b}}_{i,j}, \mathbf{C})$ measures the distance from a point $\overline{\mathbf{b}}_{i,j}$ to the plane $\mathbf{C}$.
	
	\item \textbf{Symmetry and parallelism}. Given pairs of curves that are detected to be symmetric or parallel, for simple formulation and computation, we first perform a resampling step to ensure that the two curves are represented by the same number of points. Then, the symmetry constraint is defined as:
	\begin{equation}\label{symmetry}
		E_{symmetry} = \sum_{k}((\frac{\overline{\mathbf{b}}_{i,k}+\overline{\mathbf{b}}_{j,k}}{2}-\mathbf{C}_{i,j})\cdot \textbf{n})^{2},
	\end{equation}
	where $\mathbf{C}_{i,j}=\sum_{k}(\mathbf{b}_{i,k}+\mathbf{b}_{j,k})/2$ and $\textbf{n}$ is the normalized vector of $\sum_{k}(\mathbf{b}_{i,k}-\mathbf{b}_{j,k})$.
	
	Similarly, the parallelism constraint is given by:
	\begin{equation}\label{parallel}
		E_{parallel} = \sum_{k}\|\overline{\mathbf{b}}_{i,k}-\overline{\mathbf{b}}_{j,k}+\mathbf{d}_{i,j}\|^{2},
	\end{equation}
	where $\mathbf{d}_{i,j} = \sum_{k}(\mathbf{b}_{j,k}-\mathbf{b}_{i,k})$.
	
\end{itemize}

In our implementation, we formulate the above structural regularities as soft constraints, and use Lagrange multipliers to enhance these regularities by minimizing the augmented energy function using the L-BFGS algorithm~\cite{liu1989limited}.

\subsection{Curve network completion}
\label{sec:connection}
After optimizing the curve segments, we now have regularized feature curves. Unfortunately, they are disconnected and can not be used for surface reconstruction. In this step, we propose a method to extend and close these feature curves to obtain well connected curve networks. This step is important for two purposes: 1) sharp corners of an object can be recovered after connecting several endpoints of the feature curves; 2) only closed curve networks can be used to detect surface patches for surface reconstruction.

Actually, it is not a difficult task to connect several endpoints of different feature curves if their endpoints are close to each other. Given several endpoints $\mathbf{p}_{1}, \mathbf{p}_{2}, ..., \mathbf{p}_{k}$ with corresponding tangent direction $\mathbf{t}_{1}, \mathbf{t}_{2}, ..., \mathbf{t}_{k}$, the connecting point (corner) $\mathbf{p}$ can determined by minimizing the following function,
\begin{equation}\label{anchor}
	\sum_{i=1}^{i=k}(\mathbf{p}-\mathbf{p}_{i})\cdot \mathbf{t}_{i}.
\end{equation}

\begin{figure}[t]
	\centerline
	{
		\includegraphics[width=0.7\linewidth]{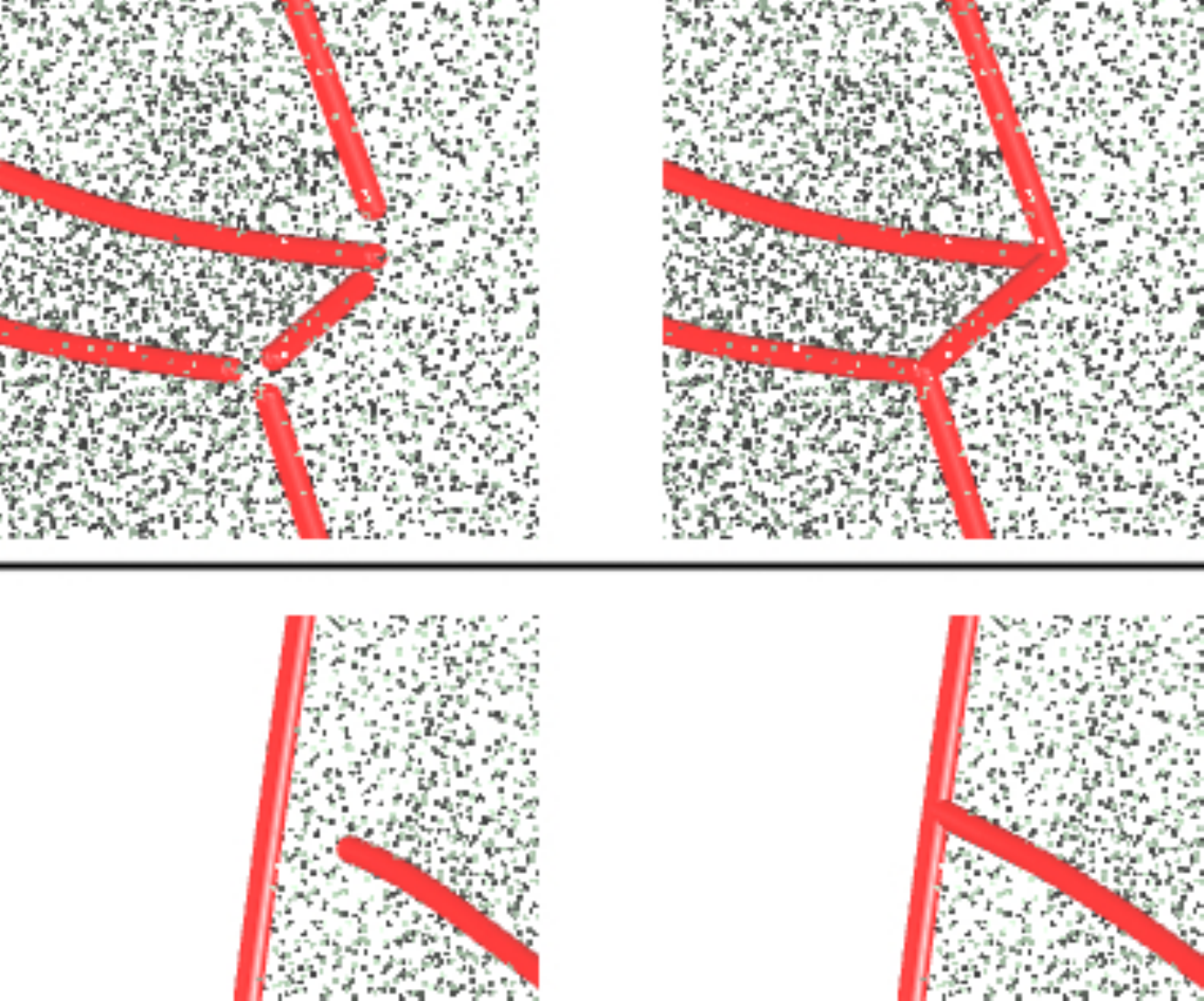}
	}
	\caption{Two types of feature curve completion. Top row: extending several feature curves to be connected at their endpoints. Bottom row: extending a feature curve to the interior of another one. Left column shows the initial curved segments, and right column are the extended feature curves.}
	\label{fig:corner_edge_connection}
\end{figure}

The top row of Fig.~\ref{fig:corner_edge_connection} shows an example for automatically connecting several endpoints of different feature curves. In order to connect an endpoint $\mathbf{p}_{1}$ of a feature curve to the interior of another curve $\mathbf{B}_{i}$, let $\mathbf{t}_{1}$ denote the tangent direction at $\mathbf{p}_{1}$, we extend $\mathbf{p}_{1}$ to the point $\mathbf{p}$ on curve $\mathbf{B}_{i}$ minimizing $(\mathbf{p}-\mathbf{p}_{1})\cdot \mathbf{t}_{1}$. The bottom row of Fig.~\ref{fig:corner_edge_connection} shows such an example.

For endpoints that are far away from each other (this is typically true for point clouds with significant missing regions), the main task for the feature curve completion problem is to determine which curves can be connected together. We solve this problem based on a clustering method with a cost function defined as below. For each endpoint $\mathbf{p}_{i}$ of a feature curve, we define the cost of connecting it with other point $\mathbf{q}_{i}$ from another curved segment as:
\begin{equation}\label{cost_endpoint}
	\begin{array}{l}
		F(\mathbf{p}_{i}, \mathbf{q}_{i})=\frac{dist(\mathbf{p}_{i}, \mathbf{q}_{i})/s_{max}}{2+\cos(\theta(\mathbf{p}_{i}, \mathbf{q}_{i}))},
	\end{array}
\end{equation}
where $dist(\mathbf{p}_{i}, \mathbf{q}_{i})$ is the distance between points $\mathbf{p}_{i}$ and $\mathbf{q}_{i}$. If the point $\mathbf{q}_{i}$ is also an endpoint, then $\theta(\mathbf{p}_{i}, \mathbf{q}_{i})$ is the angle between the tangent directions at these two points. If the point $\mathbf{q}_{i}$ is a sample point in the interior of another feature curve, then $\theta(\mathbf{p}_{i}, \mathbf{q}_{i})$ is the angle between the tangent direction of $\mathbf{p}_{i}$ and direction of the vector $\mathbf{q}_{i} - \mathbf{p}_{i}$.

\begin{figure}[t]
	\centerline
	{
		\includegraphics[width=0.7\linewidth]{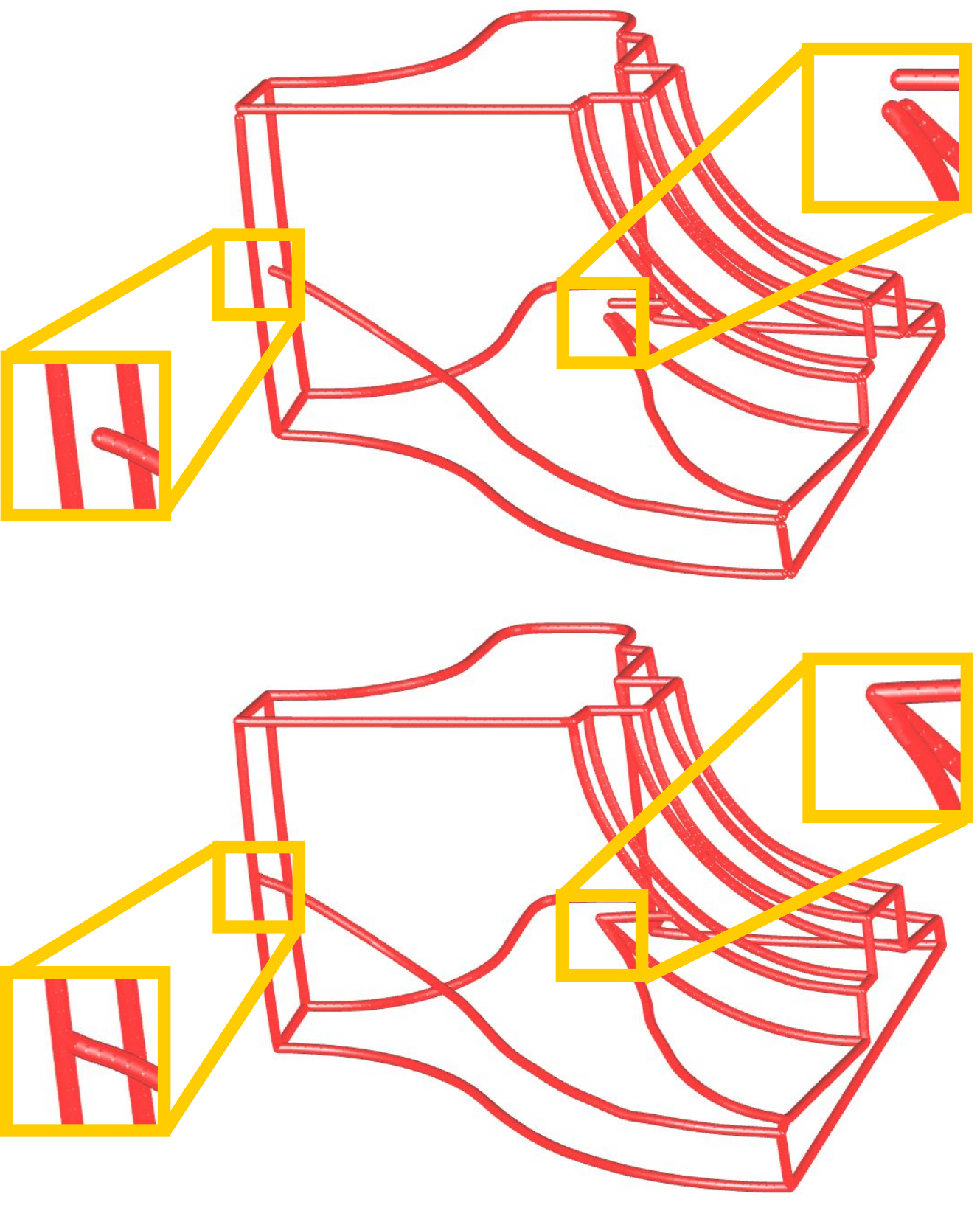}
	}
	\caption{Result of connecting several endpoints of curves}
	\label{fig:connection}
\end{figure}

For each endpoint $\mathbf{p}_{i}$, a point $\mathbf{p}_{j}$ with minimum cost $F(\mathbf{p}_{i}, \mathbf{p}_{j}) < \lambda$ from other feature curves are progressively clustered together. The threshold $\lambda$ is set to $0.9$ in our experiment. By clustering each endpoint with the best points on other curves, we get initial clusters $K_{1}, K_{2},...,K_{m}$, where $K_{i}$ is composed of points $\mathbf{p}_{i,1},\mathbf{p}_{i,2},...,\mathbf{p}_{i,n_{i}}$. We define another cost to merge two clusters such that two endpoints $\mathbf{p}_{i,s}$ in $K_{i}$ and $\mathbf{p}_{j,t}$ in $K_{j}$ are merged:
\begin{equation}\label{cost_cluster}
	\begin{array}{l}
		E(K_{i}, K_{i})=\min \limits_{\mathbf{p}_{i,s}\in K_{i}, \mathbf{p}_{j,t}\in K_{j}}F(\mathbf{p}_{i,s}, \mathbf{p}_{j,t}).
	\end{array}
\end{equation}
We iteratively merge the closest clusters as long as the cost defined by Equation~\ref{cost_cluster} is smaller than the threshold $\lambda$. We continue this process until no more cluster pairs can be merged. Fig.~\ref{fig:connection} shows one of the feature curve completion results using the proposed clustering method.

\section{Results and Discussion}
\label{sec:results}
We tested our algorithm on a large set of noisy point clouds (both real laser scans and synthetic data) with large missing regions in the neighborhood of the feature curves.

\begin{figure}[t]
	\centering
	\includegraphics[width=\linewidth] {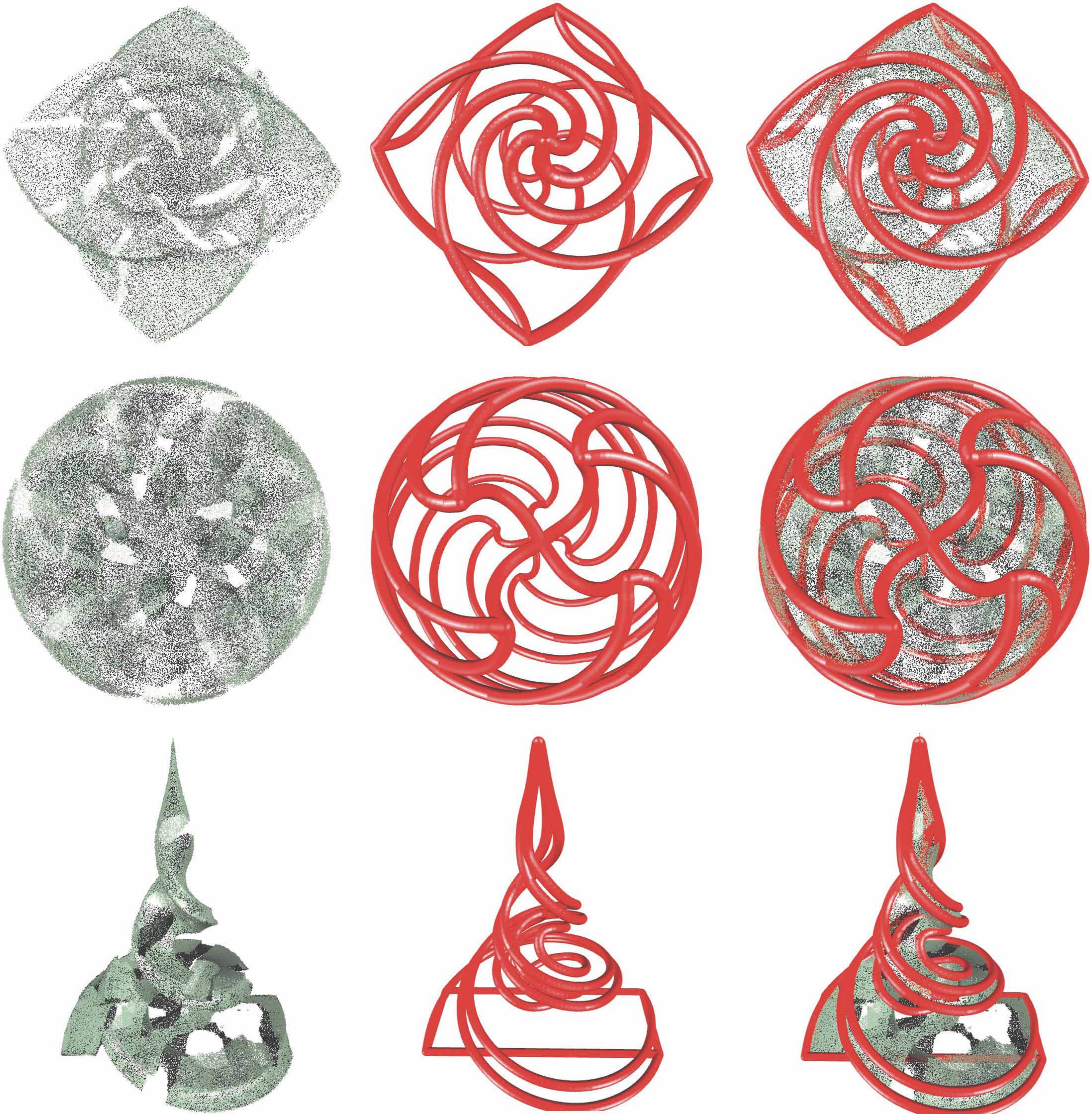}
	\caption{Curve networks extracted from point clouds of objects with free-form surfaces.}
	\label{fig:curves}
\end{figure}

\begin{figure*}[!ht]
	\centering
	\includegraphics[width=0.9\linewidth] {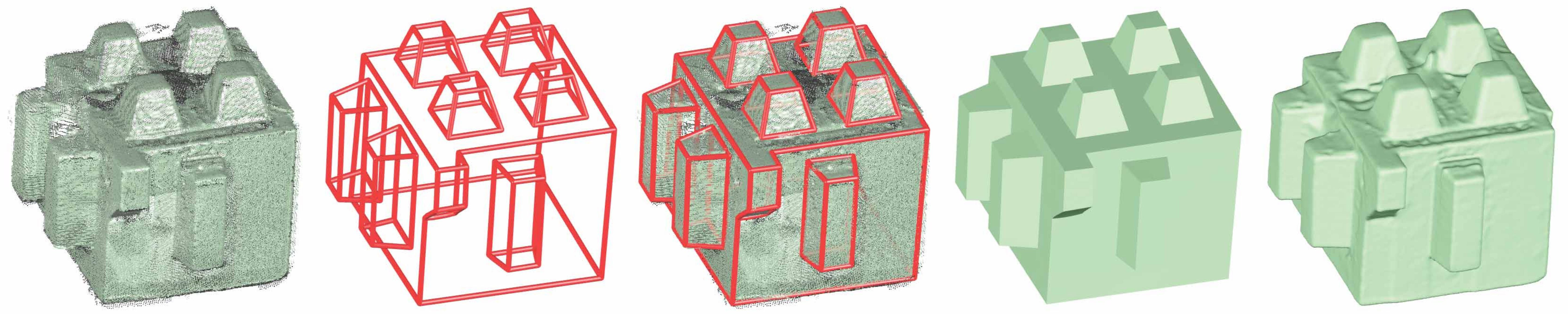}
	\caption{A mechanical part reconstructed from the curve network extracted from its noisy laser scan. From left to right: input point cloud, extracted curve network, the curve network overlaid on the point cloud, reconstructed surface model from the curve networks using the algorithm proposed by Zhuang et al.~\cite{Zhuang2013}, and the surface model reconstructed using the Screened Poisson method~\cite{Kazhdan13Poisson}. }
	\label{fig:cubes}
\end{figure*}

\begin{figure*}[!ht]
	\centering
	\includegraphics[width=0.9\linewidth] {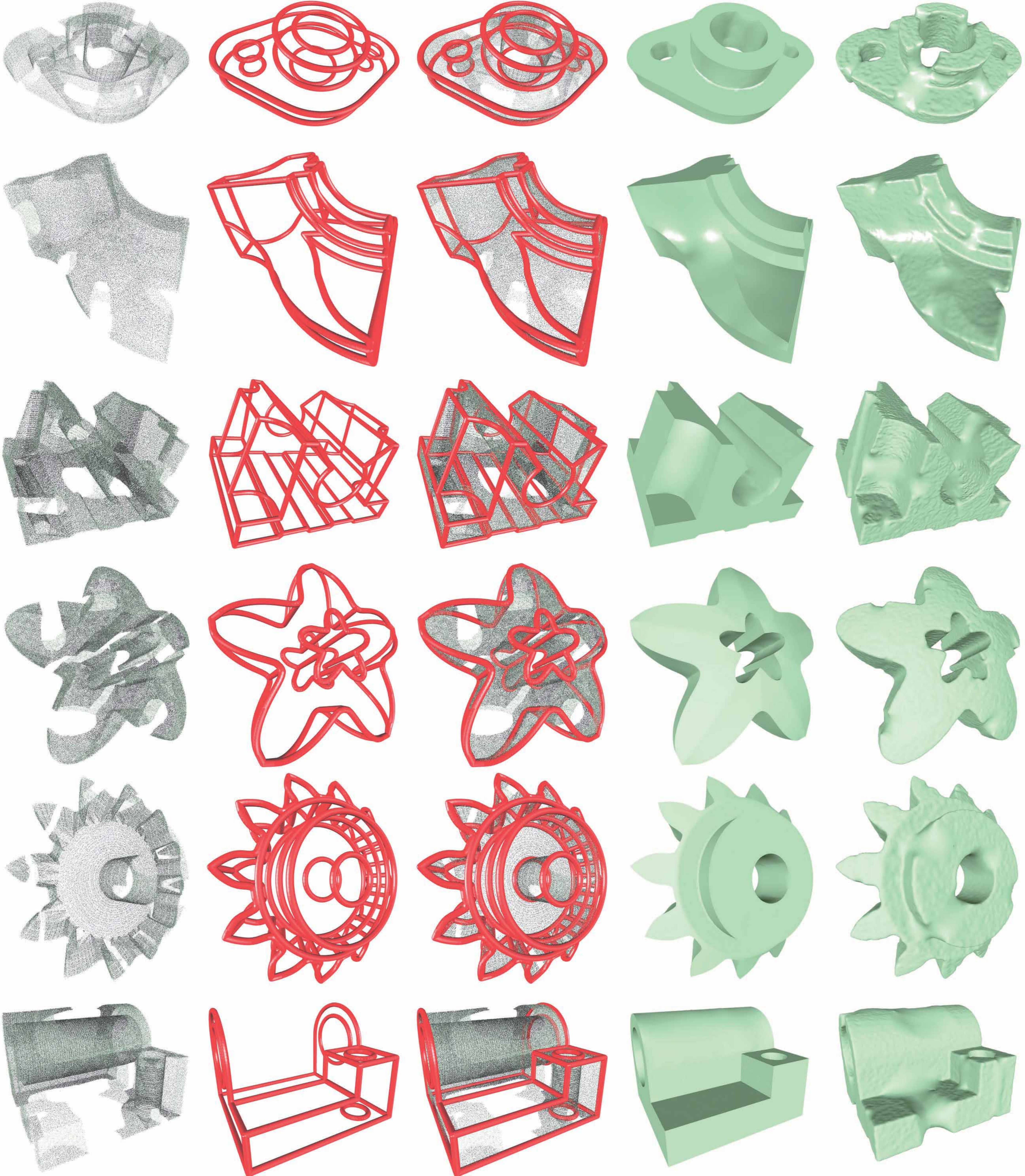}
	\caption{Curve network extraction and surface reconstruction from a set of synthetic data. From left to right: input point clouds, extracted curve networks, curve networks overlaid on the point clouds, reconstructed surface models from the curve networks using the algorithm proposed by Zhuang et al.~\cite{Zhuang2013}, surface models reconstructed using the Screened Poisson method~\cite{Kazhdan13Poisson}. }
	\label{fig:results}
\end{figure*}

\textbf{Curve network results}.
Fig.~\ref{fig:curves} shows the curve network extraction and completion results for three free-form surface objects. As can be seen from this figure, although large holes occur at the feature regions of the point clouds, our method can still produce satisfactory curve networks.

Fig.~\ref{fig:cubes} shows the processing of a mechanical part. Despite the noise in the input laser scan, our method successfully extracts the curve network from the point cloud. We can also see that the sharp features are faithfully preserved in the reconstructed surface model using our curve network as input.

In Fig.~\ref{fig:results}, we demonstrate a collection of curve networks and the corresponding surface models reconstructed from these curve networks. Similar to Figures~\ref{fig:curves} and~\ref{fig:cubes}, the examples shown here are all partial point clouds, but are mechanical parts. The first column are noisy point clouds with missing data near the feature regions. The second column are the curve networks extracted from the point clouds by our approach. The third column are the point clouds overlaid on the extracted curve networks. It can be seen clearly from this column, the extracted curve networks coincide well with the curved features of the point clouds. The fourth column shows the reconstructed surface models from the curve networks using the algorithm proposed by Zhuang et al.~\cite{Zhuang2013}.
Although some regions in the point clouds are missing (especially those at/near the feature regions), the surface models are faithfully reconstructed from the extracted curve networks.

As a comparison, we show the reconstruction results from these point clouds using the Screened Poisson reconstruction method~\cite{Kazhdan13Poisson} in the last column. It is obvious that the Poisson method can not reconstruct faithful surface models to fill large holes in the point clouds, and sharp features in the objects are usually smoothed. In contrast, the reconstruction from our extracted curve networks successfully recover these sharp features.

\textbf{Limitations}.
One limitation of our method is that it is difficult to extract very small features from the point clouds. It is also difficult to extract the feature curves that are very close to each other. In such a case, our method may not be able to separate these feature curves.

Another limitation is that we still can not handle very large missing regions. Although we developed a simple user interface to guide the feature curve completion for the regions with missing data. We found that it is still too difficult to characterize such curve features by a simple interpolation of the points.

\section{Conclusions and future work}
\label{sec:conclusion}
In this paper, we presented a framework for extracting curve networks from noisy and partial point cloud data. Our method can automatically extract and complete most of the feature curves. We exploited structural regularities to enhance the extracted curve networks to be regular and meanwhile respect the input point clouds. To resolve ambiguities for point clouds with large missing regions, we developed a simple user interface that allows the user to guide the feature extraction and completion. Experiments on various imperfect point clouds validated the effectiveness of our curve networks extraction framework. The reconstructed surface models from our curve networks confirmed that the problem of reconstruction from partial point clouds can be significantly regularized by using the curve networks extracted using our method.

In the future, we plan to exploit the extracted curve networks for further editing of the reconstructed surface models.

\section*{Acknowledgements}
\label{sec:Acknowledgements}
We gratefully acknowledge the support of NVIDIA Corporation with the donation of the Quadro K52000 GPU used for this research. This work was supported by the KAUST Visual Computing Center.

\bibliographystyle{IEEEtran}
%%use following if all content of bibtex file should be shown
%\nocite{*}
\bibliography{CurveNetworks}

\end{document}